# Atomic-scale tailoring of spin susceptibility via non-magnetic spin-orbit impurities

F.N. Womack[1], P.W. Adams[1], H. Nam[2], C.K. Shih [2] & G. Catelani [3]

Following the discovery of topological insulators, there has been a renewed interest in superconducting systems that have strong spin-orbit (SO) coupling. Here we address the fundamental question of how the spin properties of a otherwise spin-singlet superconducting ground state evolve with increasing SO impurity density. We have mapped out the Zeeman critical field phase diagram of superconducting Al films that were deposited over random Pb cluster arrays of varying density. These phase diagrams give a direct measure of the Fermi liquid spin renormalization, as well as the spin orbit scattering rate. We find that the spin renormalization is a linear function of the average Pb cluster -to- cluster separation and that this dependency can be used to tune the spin susceptibility of the Al over a surprisingly wide range from $0.8\chi_0$ to $4.0\chi_0$, where $\chi_0$ is the non-interacting Pauli susceptibility.

[1] Department of Physics and Astronomy, Louisiana State University, Baton Rouge, Louisiana 70803, USA. [2] Department of Physics, The University of Texas at Austin, Austin, TX 78712, USA. [3] JARA Institute for Quantum Information (PGI-11), Forschungszentrum Jülich, 52425 Jülich, Germany. Correspondence and requests for materials should be addressed to P.W.A. (email: adams@phys.lsu.edu)

For much of the long history of superconductivity spin-orbit effects were never at the forefront of the larger phenomenological framework. This was certainly true of development of BCS theory. Spin-orbit (SO) scattering does not break time reversal symmetry, nor does it disrupt the pairing amplitude[1]. However, it can dramatically alter the spin states of the system by destroying the spin-singlet symmetry of the ideal BCS ground state[2]. Although this was well understood by the late 1960's, the effects of spin mixing in strong SO scattering systems proved to be somewhat subtle and difficult to measure. One of its earliest reported manifestations was the Knight shift in Hg[3]. In contrast to these inauspicious beginnings, SO coupling is now believed to be a necessary component of several classes of non-conventional superconductors. These include correlated systems having non-centrosymmetric crystal structures such as $CePt_3Si$[4,5] and BiPd[6,7], as well as possible topological superconductors such as $Cu_{x-}Bi_2Se_3$[8]. The interplay between SO coupling and superconductivity is also crucial for the possible realization of Majorana fermions in proximitized nanowires[9].

Notwithstanding the resurgent interest in the SO underpinnings of non-centrosymmetric and topological superconductivity, details of how a otherwise low SO superconductor accommodates a spin-orbit impurity remains unclear[10]. This is particularly true in the case of an interacting system for which Fermi-liquid (FL) renormalizations of basic electronic properties such as the effective mass and spin susceptibility must be included. In this report, we present Zeeman-limited critical field studies of ultra-thin superconducting Al films that were grown over well-separated Pb clusters. We show that the Pb clusters not only serve as spin-orbit impurities but also have a profound effect on the $e-e$ interaction renormalization of the spin susceptibility as described in FL theory[11–13].

## Results

**Parallel critical field measurements**. The temperature dependence of the parallel (to the film surface) critical magnetic field was measured in 15 monolayer-thick superconducting Al films having varying densities of embedded Pb clusters. The clusters were typically well defined, each consisting of only a few Pb atoms. Their average separation $d$ was measured directly from an in situ high resolution scanning tunneling microscope. The thickness of the Al films used in this study was much less than superconducting coherence length, $\xi \sim 300$ Å. In this limit the orbital response to an applied parallel magnetic field is suppressed and the critical field transition is mediated by the Zeeman splitting of the conduction electrons[2]. The Zeeman-limited phase diagram gives one a direct probe of the spin properties of the superconducting condensate. If the SO scattering rate is low, as it is in pristine Al films, the low temperature first-order critical field transition is expected to be near the Clogston-Chandrasekhar[14,15] value $\mu_B H_{CC} = \Delta_0/\sqrt{2}$, where $\Delta_0 \approx 1.76 k_B T_c$ is the zero temperature gap, and $\mu_B$ is the Bohr magneton[16].

The spin properties of the BCS condensate are primarily influenced by: (1) Landau FL renormalization the spin susceptibility[13] and (2) spin-orbit scattering which inhibits spin polarization. The Zeeman critical field, itself, is also influenced by these mechanisms, as well as by the reduced film thickness $t/\xi$, where $\xi$ is the Pippard coherence length[17]. The quasi-classical theory of weak-coupling superconductivity[18,19] (QCTS), as applied to the Zeeman-limited superconductivity[20–22], captures these mechanisms via the corresponding dimensionless parameters[23]: the antisymmetric FL $G^0$, the spin-orbit $b = \hbar/(3\tau_{so}\Delta_0)$, where $\tau_{so}$ is the spin-orbit scattering time, and the orbital pair breaking $c \propto Dt^2$, where $D$ is the electron diffusivity and $t$ is the film thickness. $G^0$ is a measure of the renormalization of the spin

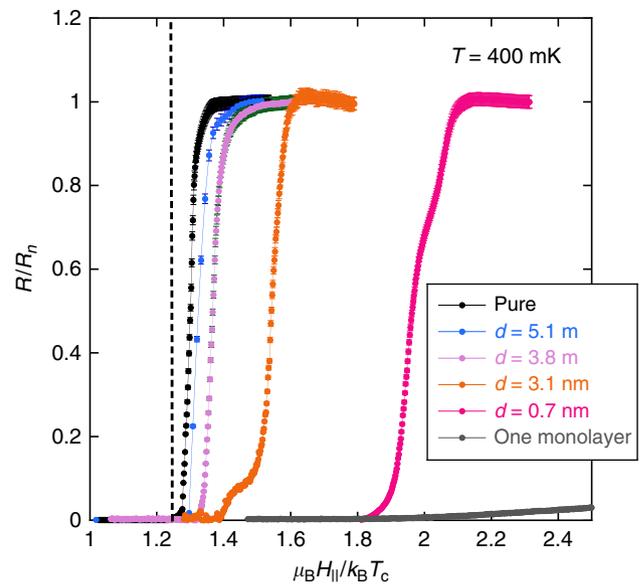

**Fig. 1** Plot of the parallel critical field transition of several 15 monolayer-thick Al films having varying Pb-cluster densities. The average Pb-cluster separation for each film is denoted by $d$. The 1 monolayer (ML) Pb trace corresponds to a 15 ML Al film deposited on 1 ML of Pb. The vertical dashed line represents the ideal Zeeman critical field. Error bars were estimated from the standard deviation of multiple measurements of the normal state resistance

susceptibility of an interacting Fermi gas, $\chi = \chi_0/(1 + G^0)$, where $\chi_0$ is the spin susceptibility of a free Fermi gas of effective mass $m^*$.

Numerous studies of the Zeeman critical field transition in ultra-thin Al and Be films have shown that these two light elements have a very low intrinsic spin-orbit scattering rate[24,25] and are true spin-singlet superconductors. Consequently, they make ideal candidates for systematic studies of the effects of SO scattering with a controllable SO impurity density. Early Zeeman critical field studies of Be and Al films showed that one could introduce SO scattering by simply coating them with sub-monolayer coverages of heavy metals (Z = Au, Pt, or Pb). These studies showed two primary effects on the critical transition. First, SO increases the Zeeman critical field well beyond the Clogston-Chandrashekar limit, due to the fact that SO scattering inhibits the polarization of the spins. Second, the presence of even modest SO scattering drives the transition from first-order to second-order[26].

The parallel critical field transitions of 15 monolayer (ML) Al films with varying cluster densities is shown in Fig. 1. The mean-free-path and coherence length of the films were determined by perpendicular critical field measurements as described in ref.[16]. The low temperature sheet resistances of films $\sim 10\,\Omega$ were insensitive to the cluster density for the range of coverages used in this study. The vertical dashed line represents the Clogston-Chandrasekhar ($H_{CC}$) critical field for the ideal case of $b = G^0 = c = 0$. The critical field of the pure Al film is slightly higher than $H_{CC}$ due to the fact that the SO and FL parameters are not exactly zero in Al. Note that the critical field increases substantially with decreasing cluster separation. We also include the critical field curve of a 15 ML Al that was deposited on 1 ML of Pb. It's critical field was $H_c \sim 8$ T, thus only the tail of critical field trace appears in the plot. In the analysis that follows we define the critical field by the midpoint of the transition.

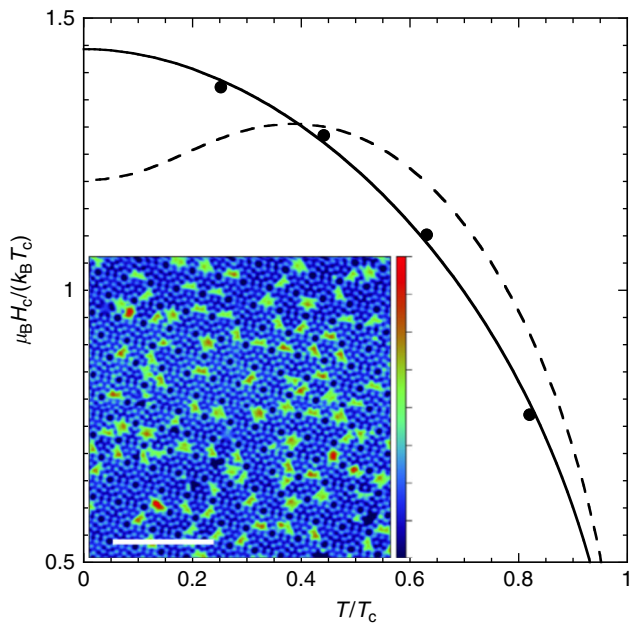

**Fig. 2** Zeeman-limited superconducting phase diagram of a 15 monolayer-thick Al film with a Pb-cluster separation of 3.8 nm. $H_c$ is the parallel critical field and $T_c$ is the superconducting transition temperature. The solid line represents a best least-squares fit to the phase diagram in which the spin-orbit parameter $b$ and the Fermi liquid parameter $G^0$ where varied. The dashed line is the corresponding fit in which only $b$ was varied with the Fermi liquid parameter set to its pristine Al film value $G^0 = 0.18$. Inset: In situ scanning tunneling microscope image of few-atom-size Pb clusters on a Si(111)-7 × 7 surface. The cluster array was subsequently covered with a 15 monolayer-thick epitaxial Al film. The horizontal bar corresponds to 10 nm. The vertical bar is the topographical height scale which varies from −20 pm (blue) to 120 pm (red)

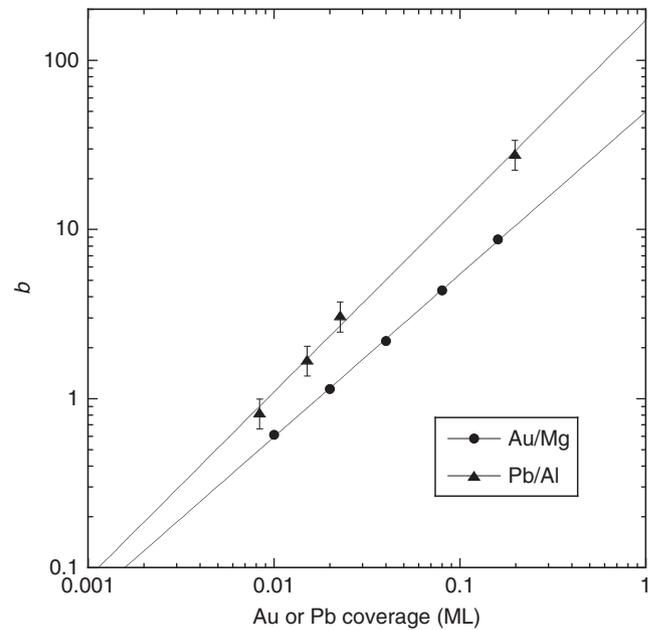

**Fig. 3** Spin-orbit parameter as a function of heavy metal coverage. The triangle symbols were obtained from quasi-classical theory of weak-coupling superconductivity fits to the Zeeman-limited superconducting phase diagrams of 15 monolayer-thick Al films as a function of the underlying Pb-cluster coverage, measured in monolayers (ML). The solid line is a power-law fit which gives an exponent of 1.1. The circle symbols represent the effective spin-orbit parameter, as determined from weak localization measurements from ref.[27] for Mg films dusted with sub-monolayer coverages of Au. The solid line through these data is a power-law fit that gives an exponent of 1.0. Error bars where estimated from standard deviations produced by the least-squares fitting algorithm

**Numerical analysis of the phase diagrams.** In Fig. 2 we plot the temperature dependence of the Zeeman critical field $H_c$ of an Al film with a Pb-cluster separation of 3.8 nm. These data represent the Zeeman-mediated phase diagram of the film. The solid line is a best least-squares fit to QCTS where only the SO parameter $b$ and the FL parameter $G^0$ were varied. The thickness parameter was previously determined from a pure 15 ML Al film. The details of the fitting procedure and its underlying assumptions has been published elsewhere[16]. Here we mention that as long as the phase transition remain second-order, as is the case for the data in Fig. 2, the critical field is obtained as solution to the equation[20]

$$\ln \frac{T}{T_c} + \tfrac{1}{2}\left(1 + b/\sqrt{b^2 - \tilde{h}_c^2}\right)\psi\left(\tfrac{1}{2} + \rho_+\right)$$
$$+ \tfrac{1}{2}\left(1 - b/\sqrt{b^2 - \tilde{h}_c^2}\right)\psi\left(\tfrac{1}{2} + \rho_-\right) - \psi\left(\tfrac{1}{2}\right) = 0, \quad (1)$$

where $\psi$ denotes the digamma function, $\tilde{h}_c = h_c/(1 + G^0)$, $h_c = \mu_B H_c/\Delta_0$, and

$$\rho_\pm = \frac{\Delta_0}{2\pi T}\left(b + ch_c^2 \pm \sqrt{b^2 - \tilde{h}_c^2}\right). \quad (2)$$

Note that this fitting procedure captures the salient features of the phase diagram. In contrast, if we fix the FL parameter to its pure Al film value $G^0 \simeq 0.18$ and only vary $b$, then the fit is much worse, as indicated by the dashed line in Fig. 2. This was also recognized in the early work of Tedrow and Meservey[26] who attempted to fit the phase diagram of Pt-coated Al films, where Pt was used to induce SO scattering. They found that for relatively large values of $b$, the measured critical fields were in poor agreement with theory, however they did not include FL corrections in their analysis.

Shown as triangle symbols in Fig. 3 are the values of the SO parameter $b$ obtained from samples of varying Pb-cluster density as a function of the cluster coverage on the Si substrate. For point-like impurities with uncorrelated positions the scattering rate, and hence $b$, is expected to be proportional to the impurity density. Therefore, $b$ should scale as the Pb coverage, which is itself proportional to $1/d^2$. The solid line represents an power-law fit to the data and gives an exponent of 1.1. We can compare our SO scattering rates with those obtained via weak localization measurements on thin Mg films dusted with sub-monolayer coverages of Au[27]. Of course, Mg films do not superconduct but we can nevertheless extract an effective $b$ for the Mg/Au data by simply multiplying the reported SO scattering rates by $\hbar/(3\Delta_0)$, where $\Delta_0$ is the average gap energy of our Al films. These data are depicted by the circle symbols in Fig. 3. The Mg/Au exhibits an exponent of 1 indicating that the SO scattering rate is simply proportional to the coverage. In our case, the Pb clusters are not point-like and their positions, while random, display some correlations over the length scale $d$. Such correlations can play an important role in the mobility in doped semiconductors and graphene and could perhaps contribute to the slightly super-linear dependence of $b$ on Pb coverage[28]. Nevertheless, the overall agreement between these two very different experimental probes of heavy element SO scattering is reassuring.

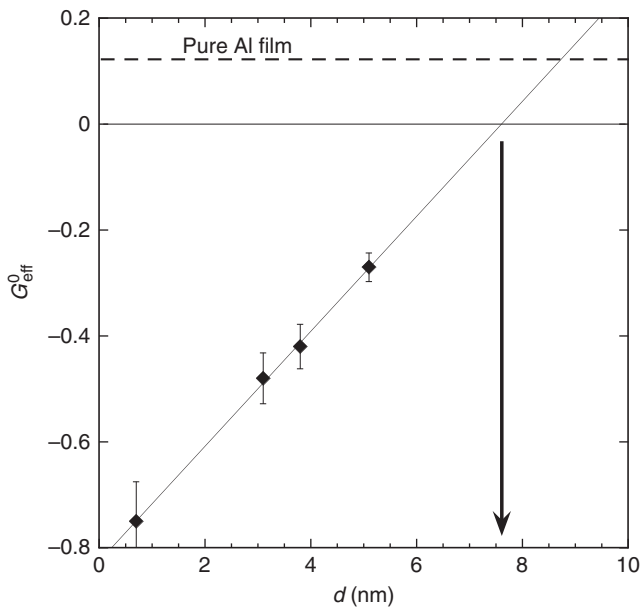

**Fig. 4** Effective antisymmetric Fermi liquid parameter as a function of Pb-cluster separation. Note that $d \sim (\text{Pb coverage})^{-\frac{1}{2}}$. The solid line is a linear least-squares fit to the data. The horizontal dashed line represents $G^0$ for a pure Al film. The arrow represents the average cluster separation at which there is no spin renormalization. Error bars where estimated from standard deviations produced by the least-squares fitting algorithm

Perhaps, the most surprising finding in these analyses is that the antisymmetric FL parameter $G^0$ is also dramatically affected by the Pb clusters. In fact, in order to fit phase diagrams like that in Fig. 2, we must treat $G^0$ as an effective free parameter, which we denote with $G^0_{\text{eff}}$ to distinguish it from that of the zero SO FL theory[13]. We should point out that increasing $G^0$ increases the theoretical critical field, which is true of $b$ as well. However their influences have somewhat different temperature dependencies[16,22]. Consequently, their relative contributions can only be de-convolved by fitting across the entire phase diagram. In Fig. 4 we plot $G^0_{\text{eff}}$ as a function of $d$. The relative magnitude of the change in $G^0_{\text{eff}}$ with decreasing $d$ is non-perturbative. Indeed, our effective approach is likely not applicable at small cluster separations, since one would expect a ferromagnetic instability at $G^0 \sim -1$. However, the analysis is sound in the perturbative limit $|G^0_{\text{eff}}| \ll 1$ and our data suggests that $G^0_{\text{eff}}$ changes sign at an average separation of $d \sim 7.5$ nm, corresponding to a Pb coverage of $4 \times 10^{-3}$ ML. Specifically, the spin correlations change from antiferromagnetic-like to ferromagnetic-like at this critical separation.

## Discussion

The origin of the shift in $G^0_{\text{eff}}$ toward ferromagnetic spin correlations is unknown[29]. It is interesting that $G^0_{\text{eff}}$ is a linear function of cluster separation and not, in contrast to $b$, a function of cluster density. The linear dependence may represent a proximity effect in which the local FL environment of Pb clusters influences the average $G^0_{\text{eff}}$ of the surrounding Al in a manner that is similar to proximity-induced exchange fields in superconducting-ferromagnet bilayers[30,31]. Unfortunately, in contrast to Al, bulk Pb is diamagnetic. Consequently there is no straightforward way to independently probe the spin susceptibility and corresponding FL environment of the Pb islands.

Another possibility is that the FL spin renormalization in the Al films is transformed from the single channel value of pristine Al to a more complex effective value in the presence of SO scattering. In a low SO FL the renormalization of the spin susceptibility by $e$–$e$ interactions only depends upon a single parameter, $G^0$. However, in the presence of generic spin-orbit couplings a more complicated relationship between spin susceptibility and the strength of the various spin-dependent interaction channels emerges[32]. It may be possible to calculate $G^0_{\text{eff}}$ using ab-initio methods, similarly to e.g., the treatment of Ni clusters magnetism in Ag[33]. Alternatively, one may be able to extract $G^0_{\text{eff}}$ from a Kondo lattice-like model. It is known that the impurity spin susceptibility in these models can be affected by the Ruderman-Kittel-Kasuya-Yosid (RKKY) interaction[34].

In summary, we have exploited the Zeeman critical field of ultra-thin superconducting Al films to investigate the evolution of their spin susceptibility as a function of imbedded Pb-island separation. This technique provides a powerful and direct probe of a spin-singlet superconductor's accommodation of local non-pair breaking SO perturbations. By varying the Pb-cluster separation the antisymmetric FL parameter $G^0$ can be tuned over a wide range $G^0_{\text{eff}} \sim 0.18 \rightarrow -0.75$ with a corresponding multifold effect on the spin susceptibility. From a practical standpoint, this allows one to adjust the spin susceptibility to a specific value for the purposes of spintronics applications. For instance, our data suggests that at a separation of ~7.5 nm $G^0_{\text{eff}} = 0$. At this impurity density the spin characteristics of the Al film are transformed into that of a non-interacting Fermi gas with modest SO scattering rate, $b \approx 0.4$.

## Methods

**Transport measurements**. The magnetotransport properties of the films were measured on a Quantum Design Physical Properties Measurement System equipped with a $^3$He probe. The base temperature of system was 400 mK. Electrical contact was made to the films using a standard 4-probe geometry and phase sensitive detection of the film resistivity. The films were carefully aligned to parallel field orientation using a custom designed mechanically actuated rotating platform fitted to the probe sample mount. After alignment, the parallel critical field was measured as a function of temperature. The cluster separation of our samples varied between $d = \infty$ for the pristine Al films and $d = 0.5$ nm for the highest Pb coverages used. The Pb clusters did not appreciably affect the transition temperature of the Al films, nor did they appreciably affect their conductivity. However, as the cluster density was increased, the SO scattering rate also increased. As expected, this produced significant higher Zeeman critical fields than is typical of pristine Al films.

**Film synthesis**. The Al-Pb-cluster samples used in this study were depositing onto carefully prepared n-doped ($n \sim 10^{-15}$ cm$^{-3}$) Si(111) substrates. The substrates were cleaned by flashing them 5 times (via Joule heating) to 1200 °C, followed by an anneal at ~550 °C for 10 min. The Pb clusters were formed by first depositing a small amount of Pb at room temperature (≪1 ML) at a chamber pressure of ~$8 \times 10^{-11}$ Torr and subsequently annealing the sample at ~200 °C for 10 min. Scanning tunneling microscope (STM) topographs were then used to determine the cluster distribution characteristics. A cluster image corresponding to 0.02 ML of Pb is shown in the inset of Fig. 2. Note that the clusters are only a few atoms in size and thus have a lateral dimension that is much smaller than either the coherence length or the film thickness. Finally, 15 ML of Al was deposited on the cluster matrix at 100 K, followed by room temperature annealing for 12 h. The upper three layers of the resulting composite film was oxidized in order to produce a protective cap. Thus the metallic thickness of the Al films used in critical field studies was approximately 12 ML (~3.2 nm)[16]. We note that because of the clustering tendencies of the Pb atoms, one needs direct STM imaging of the cluster array in order to determine the average separation. If one assumes that the Pb atoms are simply randomly distributed on the Si surface, then the average separation will be substantially underestimated.

### Data availability

The data that support the findings of this study are available from the corresponding author upon reasonable request.

### Acknowledgements

The magnetotransport measurements were performed by P.W.A. and F.N.W with the support of the U.S. Department of Energy, Office of Science, Basic Energy Sciences, under Award No. DE-FG02-07ER46420. Film fabrication and characterization was performed by H.N. and C.K.S. with support from grants ONR-N00014-14-1-0330 and NSF-DMR-1506678. The theoretical analysis was carried out by G.C. We gratefully acknowledge enlightening discussions with Ilya Vekhter, Dan Sheehy, and Anton Vorontsov.

### Author contributions

The experiments were conceived by P.W.A. and C.K.S. The magnetotransport data was collected primarily by F.N.W. The epitaxial films were grown and characterized by H.N. The theoretical analysis was provided by G.C. All authors contributed to the interpretation of the results.

### Additional information

**Competing interests:** The authors declare no competing interests.

**Reprints and permission** information is available online at http://npg.nature.com/reprintsandpermissions/

**Publisher's note:** Springer Nature remains neutral with regard to jurisdictional claims in published maps and institutional affiliations.